\documentclass[letterpaper, 10 pt, conference]{ieeeconf}  % Comment this line out if you need a4paper

\IEEEoverridecommandlockouts                              % This command is only needed if 
\usepackage{bm}

\usepackage{graphicx}
\usepackage{tabularx}
\usepackage{multirow}
\usepackage{amsmath, algorithmic, algorithm}
\usepackage{dingbat}
\usepackage[normalem]{ulem}
\usepackage{graphicx} % for images (if needed elsewhere)
\usepackage{booktabs} % for \toprule, \midrule, etc.
\usepackage{array} % for enhanced table features
\usepackage{tabularx} % for automatic column width adjustments
\usepackage{hyperref} % for clickable links in footnotes
\usepackage{footnote} % for footnotes inside tables
\usepackage{xcolor} % for colored text
\usepackage{pifont} % for checkmarks and crosses
\usepackage{multirow} % for multi-row cells
\usepackage{caption} % for better caption formatting
\usepackage{tikz}
\tikzset{every node/.style={font=\normalfont}}
\usepackage{framed}
 \usepackage{arydshln} % for dashed lines
 \usepackage{xcolor}

\usepackage{svg}
\usepackage{makecell}
\usepackage{multicol}
\usepackage{footmisc}

\usepackage{tcolorbox}
\tcbuselibrary{listingsutf8}

\makeatletter
\let\labelindent\@undefined
\makeatother
\usepackage{enumitem,amssymb}
\newlist{todolist}{itemize}{2}
\setlist[todolist]{label=$\square$}
\usepackage{pifont}

\usepackage{xcolor}
\definecolor{softgreen}{RGB}{102,187,106} % muted, non-flashy

\newcommand{\tinyscript}{\tiny}

\begin{document}

\title{LOTUSim: Multi-Domain Simulator for Marine Robotics
}
\author{
    Cédric Buche$^{1,3}$,
    Juliette Grosset$^{1,2}$,
    Hélène Lechêne$^{2}$,
    Marie Dubromel$^{1,2}$,\\
    Pierig Havez-Bodivit$^{2}$,
    Malcom Neo$^{2}$,
    Julien Prodhon$^{2}$\\[1ex]
    \small $^{1}$CROSSING IRL 2010, CNRS;
    $^{2}$Naval Group, France;
    $^{3}$IMT Atlantique
}

\maketitle

\begin{abstract}
Simulation is essential for maritime robotics, supporting operator training, mission rehearsal, and human–vehicle interaction in environments where real-world testing is costly or hazardous. Existing simulators focus primarily on autonomy systems and often lack human-in-the-loop interaction and realistic environmental physics.
This paper introduces LOTUSim, an open-source, real-time maritime simulator supporting multi-user interaction across aerial, surface, and underwater robotic systems for coordinated naval-style operations.
%, with physics-based simulation of onboard sensing for realistic mission rehearsal.
%
%
The first contribution of this work is enabling real-time interactive performance for users while ensuring scalability to large fleets operating within a shared interactive simulation environment.
Validation demonstrates robust human-in-the-loop performance, maintaining strict real-time execution and high visual fidelity while scaling to large heterogeneous maritime drone swarms.
The second contribution is a computationally efficient, Ekman-inspired layered, underwater current model that captures wind-driven, depth-dependent flow dynamics with sufficient physical fidelity for large-scale simulations. Validation against ocean reanalysis data demonstrates substantially improved accuracy compared to commonly used stochastic Gauss–Markov current models.
These results confirm LOTUSim’s suitability as a simulation platform for operator-in-the-loop maritime robotics research.

\end{abstract}

\section{Introduction}

For naval operations, human-in-the-loop (HITL) simulation is essential. Maritime missions are conducted under strong environmental forcing, limited sensing capabilities, and high cognitive workload, where operator decisions  influence mission success and safety. Interactive simulation interfaces can significantly improve situational awareness, skill acquisition, and crew coordination by allowing operators to interact naturally with vehicles and their environment \cite{Chen2025SymbioSim}. %At the same time, simulation provides a controlled yet scalable foundation for AI \cite{Song2025OceanSim}; 

Despite recent advances, existing marine robotic simulators are designed primarily for autonomous systems and sensor simulation, with limited consideration for real-time operator-centric interaction and collaborative mission execution. Even simulators actively maintained and updated in 2025, such as OceanSim \cite{Song2025OceanSim}, MarineGym \cite{marineGym}, Stonefish \cite{Grimaldi2025} and HoloOcean \cite{Potokar22icra, romrell2025previewholoocean20}, do not yet provide native support for immersive, interactive operation or multi-user mission rehearsal involving human operators in the loop.

Environmental modelling represents a second major barrier to operational realism in existing marine robotic simulators. Many simulators neglect underwater currents or use simplified models: USVsim \cite{Paravisi2019} relies on Computational Fluid Dynamics (CFD)-based precomputed fields, LRAUVSim \cite{Player2023MultiAUV} and MarineGym \cite{marineGym} uses a constant unidirectional current, and StoneFish \cite{Grimaldi2025} allows arbitrary profiles without physical realism. HoloOcean \cite{romrell2025previewholoocean20} plans to implement more realistic volumetric currents, though currently only user-defined vector fields (vortex fields) are supported.
The Gauss–Markov process, implemented in UUVSim \cite{Manhaes2016} and stratified in DAVE \cite{zhang2022dave}, provides a real-time, computationally efficient baseline. While practical, these approaches are insufficient for naval operations, where depth-dependent, wind-driven currents strongly affect vehicle behaviour, sensor performance, and operator decision-making. 

\begin{figure}[htbp]
    \centering
    \includegraphics[width=.8\linewidth]{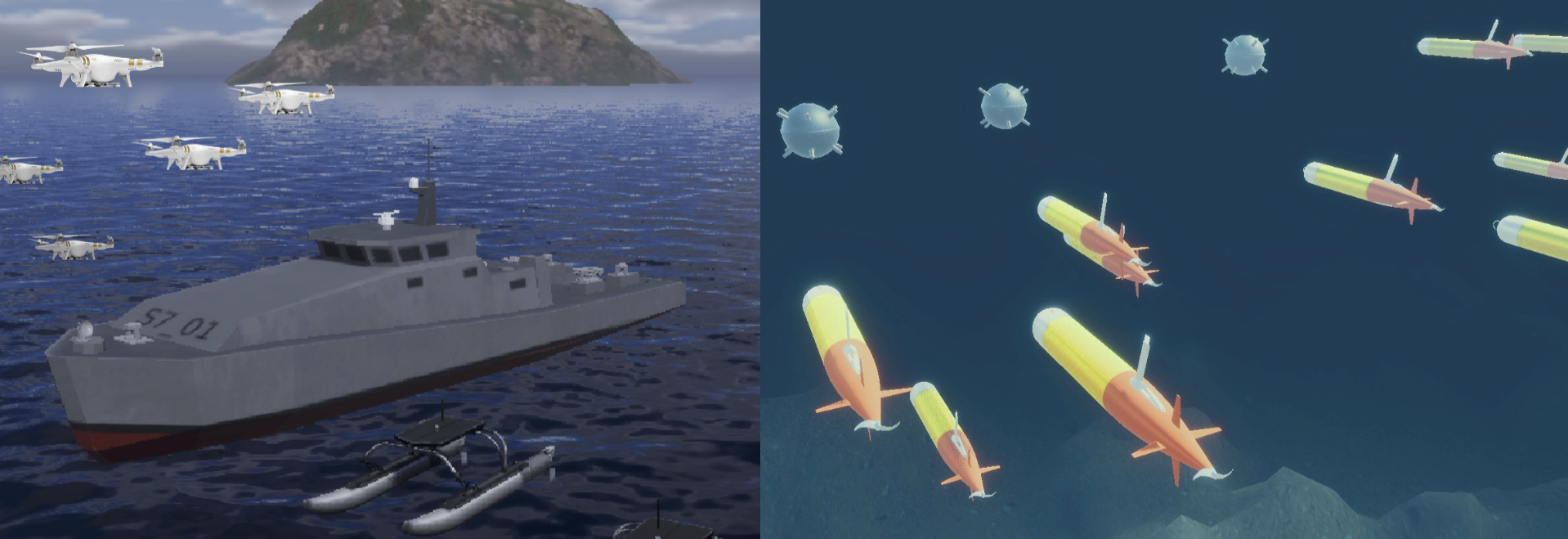}
    \caption{LOTUSim, an open multi-domain simulator}
    \label{fig:simulator}
\end{figure}

This paper presents \textbf{LOTUSim}\footnote{LOTUSim open-source code: \tiny\url{https://github.com/naval-group/LOTUSim}}$^,$\footnote{\label{fn:video}LOTUSim Video: \tiny\url{https://www.youtube.com/watch?v=iXDz8ZqSpq4}}, an open-source multi-domain maritime simulator (Figure \ref{fig:simulator}), designed to support naval-style operations and 
%digital-twin–oriented workflows 
involving both human operators and robotic systems. LOTUSim targets training, mission rehearsal in complex maritime environments, where environmental forcing, limited observability, and operator workload play a central role. To this end, the simulator enables real-time interaction between operators and heterogeneous robotic platforms across aerial, surface, and underwater domains within a distributed execution framework.

The main contributions of \textit{LOTUSim} are:
\begin{itemize}
    \item \textbf{A framework enabling real-time interaction}, demonstrating robust interactive performance and scalability for large heterogeneous fleets while maintaining real-time responsiveness.
    \item \textbf{A high-fidelity underwater current model} for real-time simulation, with an Ekman-inspired layered current model, validated against ocean reanalysis data and outperforming Gauss--Markov baseline models.
    \item \textbf{A comprehensive suite of robotic and immersive features}, integrating diverse onboard sensors (LiDAR, Radar, camera) with human-centric interfaces (Head Mounted Display, eye-tracking, Leap Motion).
\end{itemize}

\section{Review of marine robotic simulators}

\label{sec:review}

Physics simulators for maritime applications allow researchers to experiment  in a controlled environment before deploying. Additionally, simulations can be executed faster than real-time, which is particularly beneficial for learning-based approaches \cite{Player2023MultiAUV}. 
In this context, this section provides a review of maritime simulators in robotics, focusing on their ability to simulate aerial, surface, and underwater conditions. 
Although aerial, surface, and underwater environments are connected, their distinct physical properties demand different simulation models. Aerial domains involve wind and turbulence, while surface and underwater environments require modelling of buoyancy, waves, and hydrodynamic forces. To address these differences clearly, the review is divided into two parts: aerial simulations first, followed by surface and underwater domains.
%

%%%%%%%%%
\subsection{Aerial}

Table~\ref{tab:simulation_comparison} categorises various simulators according to the criteria defined in Table~\ref{tab:table1}.
In the context of a real-time maritime simulation platform for human-vehicle interaction, the expected feature is primarily real-time models for dynamic adaptation with optional interest for wind variation (time and space), obstacle interaction, boundary restrictions.
Many simulators, such as AirSim \cite{airsim2017fsr}, employ simplified wind models that do not account for time and space dependent variability.  
Gazebo \cite{Koenig2004} and Stonefish \cite{Grimaldi2025} incorporates dynamic wind modelling based on time- and space-dependent variations\footnote{\tiny\url{https://github.com/gazebosim/gz-sim/tree/gz-sim9/src/systems/wind_effects}}.  
Models based on CFD, such as OpenFOAM \cite{Weller1998}, excel in simulating wind interactions with structures (e.g., obstacles)~\cite{Kakavitsas2024}, and are adopted by simulators such as USVsim \cite{Paravisi2019}.
However, CFD has notable limitations, including its reliance on pre-computed data \cite{Kakavitsas2024}.  
In conclusion, Gazebo's or Stonefish's wind modelling offers a practical and efficient solution.

\begin{table}[htbp]
    \caption{Criteria for Aerial modelisation}
    \label{tab:table1}
    \scriptsize
    \centering
    \renewcommand{\arraystretch}{1.4}
    \begin{tabularx}{.5\textwidth}{>{\raggedright\arraybackslash}l X}
        \toprule
        \textbf{Type} & \textbf{Description}  \\
        \midrule
        Wind Dependencies  & Wind variation over time and space, enabling simulation of gusts and regional wind changes. \\

        Obstacle Interaction  & Wind behaviour around obstacles like buildings and terrain, capturing deflection and turbulence. \\
        
        Bound Restrictions  & Limits wind simulations to a predefined area, common in CFD models for controlled analysis. \\
        Pre-compute  & Precomputed wind fields (e.g., CFD) for static analysis or real-time models for dynamic adaptation. \\
        \bottomrule
    \end{tabularx}
\end{table}

\begin{table}[htbp]
    \caption{Aerial Simulation Properties}
    \label{tab:simulation_comparison}
    \centering
    \scriptsize
    \renewcommand{\arraystretch}{1.6} 
    \setlength{\tabcolsep}{4pt} 
    \begin{tabularx}{\columnwidth}{|X|c|c|c|c|c|}
        \hline
        \textbf{Properties} & AirSim & OpenFOAM & Gazebo & Stonefish & \textcolor{purple}{LOTUSim} \\ \hline
        
        Wind dependencies & \ding{55} & Space & \makecell{Time\\+ Space} & \makecell{Time\\+ Space} & \makecell{Time\\+ Space} \\ \hline
        
        Obstacle Interaction & \ding{55} & \checkmark & \ding{55} & \ding{55} & \ding{55} \\ \hline
        
        No Bound Restricted & \checkmark & \ding{55} & Optional & Optional & Optional \\ \hline
        Compute online & \checkmark & \ding{55} & \checkmark & \checkmark & \checkmark \\ \hline
    \end{tabularx}
\end{table}

%%%%%%%%%%%%%%%%

\subsection{Surface and Underwater}

Table~\ref{tab:comparison_simulators} provides a comparative overview of representative maritime simulators spanning surface and underwater domains. Recent simulators such as HoloOcean~2.0~\cite{Potokar22icra, romrell2025previewholoocean20}, Stonefish~\cite{Grimaldi2025}, MarineGym \cite{marineGym}, OceanSim~\cite{song2025oceansimgpuacceleratedunderwaterrobot}, and UNavSim~\cite{amer2023unavsimvisuallyrealisticunderwater} significantly advance the state of the art in terms of rendering quality, physics fidelity, and scalability. However, most existing platforms remain specialised either in surface or underwater operations, and mature cross-domain solutions remain limited.
Several widely used simulators, including UUVsim~\cite{Manhaes2016}, DAVE~\cite{zhang2022dave} and URSim~\cite{Katara2019}, are no longer actively maintained, which restricts their long-term applicability. While more recent simulators address maintenance, performance, and visual realism, most lack native support for immersive HITL interaction and collaborative multi-user operation. In Table~\ref{tab:comparison_simulators}, Immersive HITL refers to operator interaction through immersive interfaces such as VR, motion tracking, or embodied control, rather than conventional keyboard or graphical interfaces. Multi-user denotes the ability for multiple human operators to simultaneously interact within a simulation environment, which is distinct from multi-robot or multi-agent simulation\footnote{\tiny
~Following Ferber \cite{ferber1999multi}, a Multi-Agent System (MAS) consists of autonomous agents operating in a distributed environment with synchronized communication. In this context, MAS support refers to dynamic agent addition and removal across distributed nodes, combined with a fair and adaptive scheduler that mitigates execution-order bias, dynamically allocates computational time, enables simulation time scaling, and supports real-time operator interaction.}.

These capabilities are essential for training, mission rehearsal, and human–vehicle interaction, yet remain largely absent from current maritime simulators, including recent state-of-the-art platforms. While ROS2 support and improved physics engines are increasingly adopted, the lack of immersive and collaborative interaction highlights a gap in the literature. LOTUSim is designed to address this gap by combining cross-domain simulation, realistic environmental modelling, immersive HITL interaction, and native multi-user support within a unified and actively maintained framework.

\begin{table*}[h!]
    \centering
    \scriptsize
    \caption{Comparative overview of representative maritime simulators and supported features. 
    \scriptsize(Legend: S = Surface, UW = Underwater, GM = Gauss–Markov.)}
    \label{tab:comparison_simulators}
    \renewcommand{\arraystretch}{1.2}
    \setcellgapes{3pt}
    \renewcommand\cellalign{cc}

    \begin{tabularx}{\textwidth}{|>{\raggedright\arraybackslash}X|*{11}{>{\centering\arraybackslash}X|}}
    \hline
    & \tinyscript UUVsim
    & \tinyscript URSim
    & \tinyscript USVsim
    & \tinyscript DAVE
    & \tinyscript LRAUVSim
    & \tinyscript UNavSim
    & \tinyscript OceanSim
    & \tinyscript MarineGym
    & \tinyscript StoneFish
    & \tinyscript HoloOcean
    & \tinyscript \textcolor{purple}{LOTUSim} \\
    \hline

    \tinyscript \textbf{Reference}
    & \cite{Manhaes2016} & \cite{Katara2019} & \cite{Paravisi2019} & \cite{zhang2022dave}
    & \cite{Player2023MultiAUV} & \cite{amer2023unavsimvisuallyrealisticunderwater} 
    & \cite{song2025oceansimgpuacceleratedunderwaterrobot} & \cite{marineGym} 
    & \cite{Grimaldi2025} & \cite{Potokar22icra, romrell2025previewholoocean20} 
    & / \\
    \hline

    \tinyscript \textbf{Domain}
    & UW & UW & S & UW & UW & UW & UW & UW & S+UW & S+UW & S+UW \\
    \hline

    \tinyscript \textbf{Launching}
    & 2016 & 2019 & 2019 & 2022 & 2023 & 2023 & 2025 & 2025 & 2019 & 2022 & 2025 \\
    \hline

    \tinyscript \textbf{Maintained}
    & \ding{55} & \ding{55} & \ding{55} & \ding{55} & \textcolor{softgreen}{\checkmark}
    & \textcolor{softgreen}{\checkmark} & \textcolor{softgreen}{\checkmark} & \textcolor{softgreen}{\checkmark} 
    & \textcolor{softgreen}{\checkmark} & \textcolor{softgreen}{\checkmark} & \textcolor{softgreen}{\checkmark} \\
    \hline

    \tinyscript \textbf{ROS2}
    & \ding{55} & \ding{55} & \ding{55} & \ding{55} & \ding{55} 
    & \textcolor{softgreen}{\checkmark} & \textcolor{softgreen}{\checkmark} & \ding{55} 
    & \textcolor{softgreen}{\checkmark} & \textcolor{softgreen}{\checkmark} & \textcolor{softgreen}{\checkmark} \\
    \hline

    \tinyscript \textbf{MAS}
    & \textcolor{softgreen}{\checkmark} & \ding{55} & \ding{55} & \textcolor{softgreen}{\checkmark} 
    & \textcolor{softgreen}{\checkmark} & \ding{55} & \ding{55} & \ding{55} 
    & \textcolor{softgreen}{\checkmark} & \textcolor{softgreen}{\checkmark} & \textcolor{softgreen}{\checkmark} \\
    \hline

    \tinyscript \textbf{Physics server}
    & Gazebo & Unity & Gazebo & Gazebo & Gazebo & AirSim 
    & \makecell[c]{Isaac Sim} & \makecell[c]{Isaac Sim} & Bullet & Unreal Engine & Gazebo \\
    \hline

    \tinyscript \textbf{Rendering}
    & Gazebo & Unity & Gazebo & Gazebo & Gazebo & Unreal Engine 
    & \makecell[c]{Omniverse\\ RTX} & \makecell[c]{Omniverse\\ RTX} & OpenGL & Unreal Engine & Unity \\
    \hline

    \textcolor{purple}{\tinyscript \textbf{Multi-user}}
    & \ding{55} & \ding{55} & \ding{55} & \ding{55} & \ding{55} 
    & \ding{55} & \ding{55} & \ding{55} & \ding{55} & \ding{55} & \textcolor{softgreen}{\checkmark} \\
    \hline

    \tinyscript \textcolor{purple}{\textbf{Immersive HITL}}
    & \ding{55} & \ding{55} & \ding{55} & \ding{55} & \ding{55} 
    & \ding{55} & \ding{55} & \ding{55} & \ding{55} & \ding{55} & \textcolor{softgreen}{\checkmark} \\
    \hline

    \tinyscript \textcolor{purple}{\textbf{\makecell[l]{Ocean\\current}}}
    & GM 
    & \ding{55} 
    & CFD nonuniform currents 
    & GM and stratified 
    & Constant unidirectional force 
    & \ding{55} 
    & \ding{55} 
    & Constant unidirectional force 
    & Uniform, Jet and Pipe
    & Vortex current field
    & Ekman layered model\\
    \hline

    \end{tabularx}
\end{table*}

\section{LOTUSim Architecture}
\label{sec:LOTUSim}

\subsection{Overview}

\begin{figure}[htbp]
    \centering    
    \captionsetup{justification=centering}
    \includegraphics[width=0.95\linewidth]{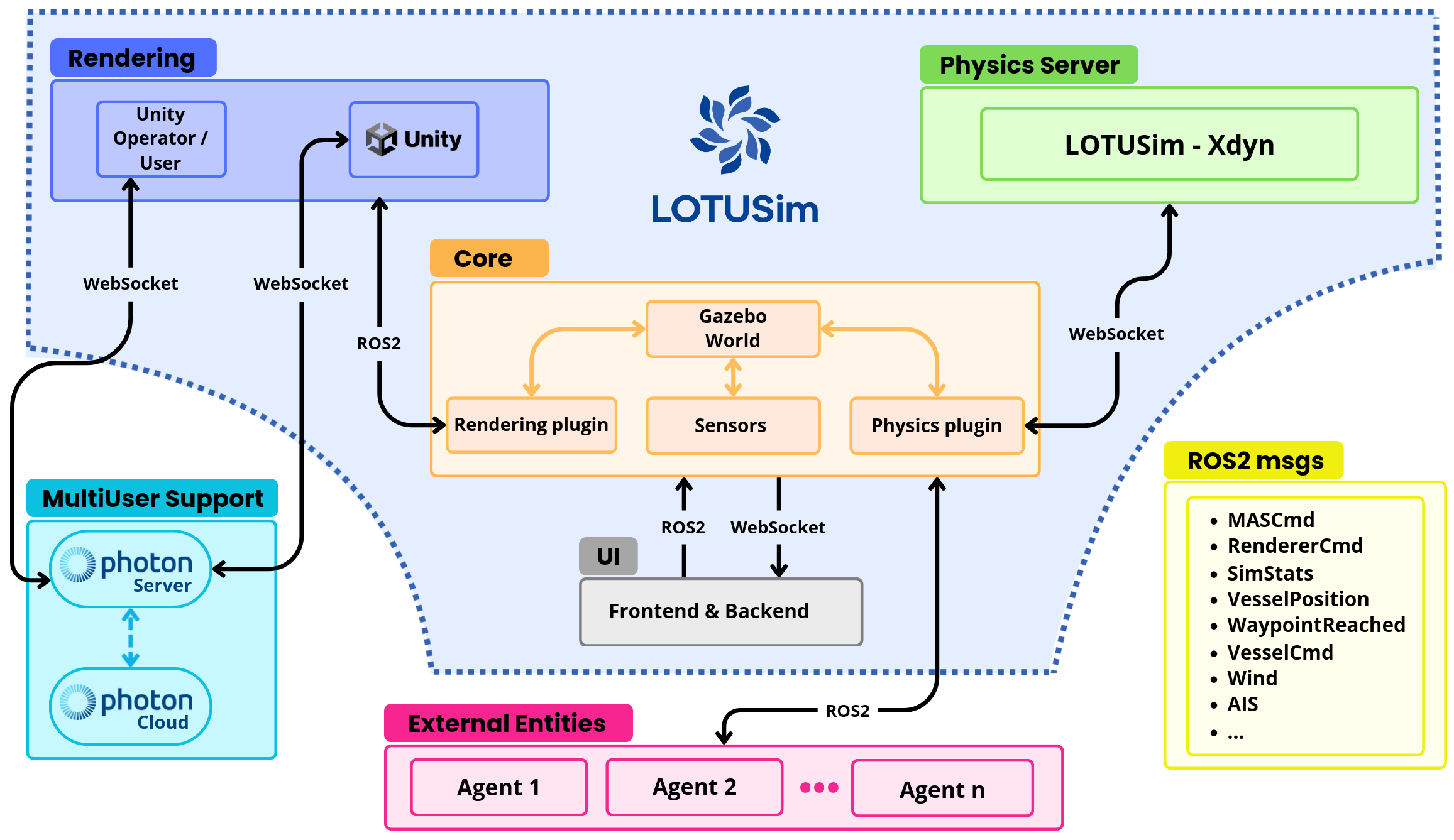}
    \caption{System architecture of LOTUSim}
    \label{fig:detailed_archi_lotusim}
\end{figure}

LOTUSim leverages ROS2, Gazebo, and Unity as \mbox{solutions} for a distributed multi-agent simulation (Figure~\ref{fig:detailed_archi_lotusim}). 
ROS2 manages the multi-agent communication and helps integrate with real drones through customised plugins, while Gazebo provides realistic physics simulation. Unity is integrated as the primary visualisation and interaction engine.
The primary goal of LOTUSim is to provide a distributed server-client-based simulation framework applicable across multiple domains, with a particular emphasis on marine use cases. Gazebo serves as the central orchestrator for asset management, while various interfaces and client modules are used to execute specific simulation models on demand.
The core module (multi-agent simulation control) is interfaced with 
three other client modules:
1/ Physics computation
2/ Rendering
3/ Agent Interaction.
Each interface supports different communication protocols. Currently, LOTUSim supports a wide range of protocols, including ROS2, WebSocket, and TCP/IP.

When using Gazebo as the asset orchestrator, each of the main client types is associated with a dedicated Gazebo plugin responsible for querying the corresponding client to update the asset state. During each simulation update cycle in Gazebo, where the timestep length is user-configurable, each asset may issue a query to its respective client module, depending on simulation requirements. The client processes the request and returns the computed result. This architecture enables computational workloads to be offloaded to external machines. The update loop concludes once all client responses are received, ensuring synchronisation across the distributed system.

\subsection{Technical Details}

\textbf{Physics computation:}
LOTUSim offloads physics calculations by transmitting each asset’s position, velocity, and the current timestep duration to an external physics engine. Since different physics engines require varying data formats and parameters, we developed a base physics interface that provides a standardised method for extracting all relevant asset information. This abstraction simplifies integration with a wide range of external physics engines.
To mitigate scheduling bias, caused by Gazebo’s default behavior of updating assets in their load order, the physics interface randomises the update order of assets in each simulation loop. Additionally, due to LOTUSim’s distributed architecture, where users may also be located on remote machines, ship control commands are transmitted via communication protocols (e.g., ROS2). This communication latency further randomises the timing of physics updates, promoting a more discrete and non-deterministic simulation environment.

Xdyn\footnote{Xdyn: \tiny\url{https://gitlab.com/sirehna_naval_group/sirehna/xdyn}} is an open-source lightweight ship simulator that models real-time vessel dynamics at sea.
For surface ships and underwater drones, we introduce an extension named as LOTUSim-Xdyn\footnote{LOTUSim-Xdyn: \tiny\url{https://github.com/naval-group/LOTUSim-Xdyn}}, detailed in Section~\ref{sec:hydro}, which connects to the simulation via WebSocket.  

%\vspace{0.08cm}
\textbf{Agent interaction:}
Currently interfaced by the ROS2 system, agent interaction can be implemented in various programming languages by communicating using ROS2 DDS with a system through different action servers and topics. 

%\vspace{0.08cm}
\textbf{Rendering and multi-user interaction:}
Visualisation is decoupled from the core simulation logic to support headless execution for AI training. When enabled, users can choose what to render and which renderer to integrate, but LOTUSim uses Unity as its default renderer. While engines like Unreal offer superior photorealism, Unity was selected for its superior extensibility and native support for immersive interfaces (e.g., VR and Leap Motion). In each timestep, the plugin will send out the locations of all assets, the newly added and destroyed assets are also published. Users are able to add custom rendering processes (e.g. explosion) into the visualisation system.

To facilitate shared missions, LOTUSim incorporates Photon Unity Networking (PUN2). This enables distributed state synchronisation, allowing multiple operators in different locations to interact with the same agents in real-time. By leveraging a cloud-relay architecture, PUN2 ensures consistent world-state replication across all Unity clients.

\subsection{Multi-Agent Architecture} 
\label{sec:mas}

The multi-agent approach improves robustness and flexibility by allowing each entity to operate independently with local decision-making.
Many works have extended Gazebo for multi-agent system (MAS) research by integrating with ROS-based frameworks \cite{ko2020ros2}.
Its flexible plugin architecture supports scalable MAS prototyping in applications such as cooperative exploration, formation control, and multi-robot task allocation.

% \subsection{Proposal}
The architecture is designed to support a flexible and distributed MAS. Each machine in the network can host one or more agents, and agents can be dynamically added or removed at runtime.
The MAS plugin serves as a centralised controller within Gazebo. It hosts a ROS2 action server that supports both bulk and individual spawning of vessels. Users can interact with this server either through the LOTUSim UI or by developing custom scripts.
During each simulation update loop, the system processes user-specified commands to spawn, delete, or move vessels accordingly. Similar to the physics update mechanism, execution order and timing are inherently randomised due to network communication delays, ensuring nondeterministic behavior and robustness in distributed environments. 

\subsection{Evaluation}

The evaluation of LOTUSim focuses on its capacity for HITL interaction, specifically measuring how the system maintains responsiveness and visual fidelity as the complexity of the swarm increases.

\textbf{Metrics for Interactivity:} We evaluate the simulator through four key metrics that directly impact human perception and control precision:

\begin{itemize}
    \item \textit{Visual Frame Rate} - Frames Per Second (\texttt{FPS}) \cite{FPSStandard} assess visual continuity. For high-stakes HITL tasks, maintaining high FPS is vital to prevent operator spatial disorientation.
    \item \textit{Physics Update Parameters} (\texttt{update\_rate}) \cite{Player2023MultiAUV, moss2010latency} monitors the physics engine’s heartbeat. A stable update rate ensures that vehicle physics react to human inputs without lag.
    \item \textit{Real-Time Factor} (\texttt{RTF}) \cite{dumnich2023lag} is the ratio of simulation time to real-time. A consistent \texttt{RTF(=1)} is required for meaningful human intervention.
    % \item Agent Scalability (number of agents): Identifies the limit at which the human operator begins to experience a degradation in system responsiveness.
\end{itemize}

\textbf{Results for Real-Time Interactive Performance:} Using the configuration in Table \ref{tab:benchmark_hardware}, LOTUSim demonstrates a robust operational envelope for complex maritime scenarios, as summarised in Table~\ref{tab:lotusim_results}.

\begin{itemize}
    \item \textit{HITL Responsiveness}: To evaluate interactive control, we measured the 
    $\overline{\texttt{update\_rate}}$  using 200 ms as the threshold for perceptually responsive operation. LOTUSim remained within this limit while scaling up to 750 Long-Range Autonomous Underwater Vehicles (LRAUVs) or 450 BlueROVs.

    \item \textit{Physics Latency}: At a 30 ms control loop (the threshold update limit), LOTUSim can simulate swarms of up to 30 BlueROV or 55 LRAUV agents while keeping physics updates strictly synchronised with real-time.

    \item Visual Rendering: Across the same configurations, LOTUSim consistently maintained a high rendering rate ($\overline{\texttt{FPS}} > 140$), providing smooth and immersive visual feedback for operators.
    
    \item Large-Scale Heterogeneous Scenarios: Under strict real-time constraints (\texttt{RTF=1}), the simulator successfully manages a large-scale heterogeneous fleet, including heterogeneous surface, underwater and aerial vessels.
    %(one highly detailed DTMB surface ship, three commando ships, and three WAMV ships), underwater vehicles (45 LRAUVs and 45 BlueROVs), and aerial drones (90 X500s).
\end{itemize}

\begin{table}[htbp]
    \centering
    \scriptsize
    \caption{Benchmark system specifications}
    \label{tab:benchmark_hardware}
    \begin{tabular}{|c|c|}
        \hline
        \textbf{Specification} & \textbf{LOTUSim} \\
        \hline
        OS - Ubuntu & 22.04.5 LTS \\
        \hline
        Processor & i9-13980HX \\
        \hline
        CPU & Up to 5.6 GHz \\
        \hline
        GPU NVIDIA GeForce RTX & 4090 Laptop \\
        \hline
        GPU VRAM & 16 GB \\
        \hline
        NVIDIA Driver & 570.133.07 \\
        \hline
        CUDA & 12.8 \\
        \hline
        Python & 3.10.12 \\
        \hline
        Gazebo & Harmonic \\
        \hline
        ROS & ROS2 Humble \\
        \hline
    \end{tabular}
\end{table}

\begin{table}[htbp]
    \centering
    \scriptsize
    \caption{Summary of Real-Time Interactive Performance}
    \label{tab:lotusim_results}
    \renewcommand{\arraystretch}{1.4}
    
    \begin{tabular}{|p{2.6cm}|p{5.2cm}|}
        \hline
        \textbf{Evaluation Aspect} & \textbf{Observed Performance} \\
        \hline
        
        HITL Responsiveness &
        750 LRAUVs or 450 BlueROVs \\
        \hline
        
        Physics Latency &
        30 BlueROVs or 55 LRAUVs \\
        \hline
        
        Visual Rendering &
        $\overline{\texttt{FPS}} > 140$ \\
        \hline
        
        Large-Scale Heterogeneous Scenario &
        \begin{tabular}[t]{@{}l@{}}
            Surface: 1 DTMB, 3 commando boats, 3 WAMVs \\
            Underwater: 45 LRAUVs, 45 BlueROVs \\
            Aerial: 90 X500s
        \end{tabular}\\
        \hline
    \end{tabular}
\end{table}

\section{Environment modelling}\label{sec:hydro}

Environmental effects such as wind, waves, and underwater currents are often modeled using real data or simplified formulations. However, real data lacks flexibility in the context of real-time or scenario-driven simulations. LOTUSim is a multi-domain maritime simulator that jointly models surface, underwater, and aerial dynamics, enabling realistic cross-domain interactions. It leverages Gazebo’s wind modeling and introduces custom surface and underwater models, including a layered ocean current model inspired by Ekman layer theory \cite{Hunkins1966,POND1983}.

\subsection{Surface}

LOTUSim-Xdyn computes ship motion using Fossen’s equations of motion \cite{Fossen2011} while incorporating detailed hydrodynamic forces, including Froude-Krylov forces and diffraction forces. It also features customisable maneuvering models and actuators. Some recent developments have been made to wind propulsion models for cargo ships \cite{Babarit2024}.

For wave simulation, LOTUSim-Xdyn uses Airy wave theory, a linear mathematical model that describes gravity wave propagation on the ocean surface \cite{Goda2010,Rodenbusch1986}. This theory assumes uniform water depth and homogeneous fluid properties, providing an efficient foundation for wave-structure interaction calculations. LOTUSim-Xdyn allows users to export environment data as 2D or 3D grids for post-analysis, including surface elevation.

\subsection{Underwater}

Several simulators adopt alternative strategies to represent ocean currents, each with specific limitations (see Table~\ref{tab:comparison_simulators}). USVsim \cite{Paravisi2019} relies on CFD-based precomputed fields, which prevent real-time simulation and limit dynamic adaptation. LRAUVSim \cite{Player2023MultiAUV} and MarineGym \cite{marineGym} implements a constant, unidirectional current, offering computational simplicity but neglecting realistic temporal or spatial variations. In contrast, StoneFish \cite{Grimaldi2025} supports completely arbitrary current profiles (uniform, jet, or pipe flows), providing flexibility for testing but lacking physical realism. Efforts such as HoloOcean~\cite{romrell2025previewholoocean20} aim to introduce volumetric effects, but ocean currents are currently implemented using user-defined vector (e.g., vortex) fields.
A widely used computational baseline is the Gauss–Markov (GM) process, which, despite lacking a physical foundation, is computationally efficient and suitable for real-time simulation. It serves as the standard real-time current model in UUVSim \cite{Manhaes2016} and DAVE \cite{zhang2022dave}, making it a convenient baseline for comparisons.

The Ekman layer framework \cite{stewart2009introduction,hautala2020physics} offers a practical method for simulating ocean currents. By combining wind measurements with classical Ekman theory, it delivers computationally efficient current estimates suitable for operational use \cite{Constantin2019,CushmanRoisin2011Introduction}. In our implementation, the water column is divided into three vertical zones under the assumption of steady-state flow. At the surface, wind stress generates currents that, under the influence of the Coriolis force, form the well-known Ekman spiral. Near the seabed, friction gives rise to a bottom Ekman layer, also exhibiting spiral patterns, while variations in bathymetry drive vertical motions. Between these layers lies the geostrophic interior, where the flow is largely in geostrophic balance and  unaffected by surface or bottom friction.

Within LOTUSim-Xdyn, we incorporated two current representations: the standard constant-current model and the new Ekman-inspired formulation described here. 

%%%%

\vspace{0.1cm}
\textbf{Parameters:}
\begin{itemize}
    \item Coriolis parameter: $f = 2 \Omega \sin(\varphi)$, where $\varphi$ is the latitude and $\Omega$ is Earth's rotation rate. 
    \item Drag coefficient: We use a simple surface drag coefficient $C_D$ inspired by Curcic and Haus \cite{Curcic2020}:
    \begin{equation}
    \scriptsize
        C_D =
        \begin{cases}
            (0.79 + 0.08\, U_{10}) \times 10^{-3}, & U_{10} < 20.5~\mathrm{m/s},\\[2pt]
            2.43 \times 10^{-3}, & U_{10} \ge 20.5~\mathrm{m/s}.
        \end{cases}
    \end{equation} 
    \item Wind stress $\tau_s = C_D \rho_{\text{air}} U_{10}^{2}$ uses the $10$\ m wind $U_{10}$. 
\end{itemize}
Note that $u,v,w$ are functions of $x,y,z$ and $t$.

%%%%

\vspace{0.1cm}
\textbf{Top layer:}

Surface currents combine Airy wave orbital velocities with the Ekman spiral (northern hemisphere formulation):
\begin{equation} \label{eq: Article_Ekman_Top}
    \scriptsize
\begin{split}      
&u = \bar{u} + V_0 e^{- \pi z_s / D_s} \cos\left(\frac{\pi}{4} - \frac{\pi z_s}{D_s} - \phi \right) \\ 
&v = \bar{v} + V_0 e^{- \pi z_s / D_s} \sin\left(\frac{\pi}{4} - \frac{\pi z_s}{D_s}  + \phi \right) \\
&w = 0
\end{split}
\end{equation}

where: \begin{itemize}
    \item $z_s$ is the depth, null at the surface and positive downwards
    \item $D_s$ is the surface Ekman layer depth
    \item $V_0$ is the current velocity at the surface
    \item $\phi$ is the wind orientation at the surface
\end{itemize}
% \cite{stewart2009introduction}

%%%%
\vspace{0.1cm}
\textbf{Middle Layer:}
\begin{equation} \label{eq: Article_Ekman_Middle}
% \begin{split}      
% &u = \bar{u} \\ 
% &v = \bar{v}  \\
% &w = 0
% \end{split}
u = \bar{u}, \quad v = \bar{v}, \quad w = 0
\end{equation}
The middle layer is unaffected by the waves or the seabed.

\vspace{0.1cm}
\textbf{Bottom Layer:}

For non-flat seabeds, correcting terms are introduced to preserve continuity and boundary conditions. In the northern hemisphere, the bottom Ekman spiral is given by
\begin{equation} \label{eq: Article_Ekman_Bottom}
    \scriptsize
\begin{split}
&u = \bar{u}\left[1 - e^{- \pi z_b / D_b}\cos\left(\frac{\pi z_b}{D_b}\right)\right] - \bar{v}e^{- \pi z_b / D_b}\sin\left(\frac{\pi z_b}{D_b}\right) \\
&v = \bar{u} e^{- \pi z_b / D_b}\sin\left(\frac{\pi z_b}{D_b}\right) + \bar{v}\left[1 - e^{- \pi z_b / D_b}\cos\left(\frac{\pi z_b}{D_b}\right)\right] \\
&w = 0
\end{split}
\end{equation}
where: \begin{itemize}
    \item $z_b$ is the depth, null at the seabed and positive upwards
    \item $D_b$ is the bottom Ekman layer depth
\end{itemize}

\subsection{Evaluation}

This section assesses the effectiveness of the underwater current models implemented within LOTUSim.

\textbf{Methodology and Setup:} The evaluation involves a statistical comparison between Copernicus in situ measurements (real-world data), the Gauss–Markov current model (UUVsim, DAVE), and the Ekman layer–based model (LOTUSim).

\begin{itemize}
    \item \textit{Study Area}: The analysis focuses on waters off the coast of Brest, France (longitude: [$-6.25^\circ$, $-6^\circ$], latitude: [$46.6^\circ$, $47^\circ$]), chosen due to its dynamic underwater currents and the availability of high-quality observational datasets.
    
    \item \textit{Depth Coverage}: Observational data span from 0.5 meters down to 1000 meters, capturing the range most relevant to surface and near-surface current dynamics.
    
    \item \textit{Temporal Sampling}: Measurements were recorded at four times per day (00:00, 06:00, 12:00, and 18:00) over five separate days throughout the year. Each day represents different wind and environmental conditions, ensuring a diverse set of oceanographic scenarios. These selected days were then used to forecast conditions for the following day (see Table \ref{tab:global_stats} for details).

\end{itemize}

\begin{table}[h!]
    \centering
    \scriptsize
    \caption{Data variability}
    \label{tab:global_stats}
    \begin{tabular}{lccc}
    \toprule
    Variable & Minimum & Maximum & Mean \\
    \midrule
    $u$ (zonal current) & -0.52 & 0.55 & 0.004 \\
    $v$ (meridional current) & -0.49 & 0.33 & 0.005 \\
    $V_\text{north}$ (northward wind) & -12.6 & 14.9 & -2.7 \\
    $V_\text{east}$ (eastward wind) & -9.6 & 22.3 & 1.2 \\
    \bottomrule
    \end{tabular}
\end{table}

\textbf{Dataset}: The analysis utilises a total of 2,800 measurement points (N) distributed across the study area and depth range, with 700 points corresponding to each observation time. For model validation, data from the following day were employed as a reference to assess predictions generated by both the Gauss–Markov model and the Ekman layer model (LOTUSim).

\textbf{Evaluation Metrics:} Model accuracy was quantified using both absolute and relative error measures for the zonal (\textit{u}) and meridional (\textit{v}) current components. Specifically, the mean absolute error (MAE) and root mean square error (RMSE) were employed, as these are widely adopted in ocean current forecasting studies \cite{bayindir_predicting_2023, Sinha2021, Menaka2021}. Definitions for these metrics are provided in Table~\ref{tab:error_metrics}. These indicators enable a thorough comparison of model performance across different depths, observation times, and varying environmental conditions.

\begin{table*}[htbp]
    \vspace*{1mm}
    \centering
    \caption{Metrics for model comparison}
    \label{tab:error_metrics}
    \scriptsize
    \renewcommand{\arraystretch}{1.4} % extra vertical padding
    \begin{tabularx}{.99\textwidth}{|>{\raggedright\arraybackslash}m{0.28\textwidth}|
                                          >{\centering\arraybackslash}p{0.25\textwidth}|
                                          >{\raggedright\arraybackslash}m{0.389\textwidth}|}
        \hline
        \textbf{Metric} & \textbf{Definition / Formula} & \textbf{Interpretation / Key Value} \\
        \hline
        \textit{Ekman vs. Gauss Ratio (MAE)} ($I_{\rm MAE}$) &
        $I_{\rm MAE} = \dfrac{\mathrm{MAE}_{\rm Ekman}}{\mathrm{MAE}_{\rm Gauss}}$ &
        Values $<1$ indicate the Ekman model outperforms the Gauss–Markov model; lower is better \\
        \hline
        \textit{Ekman vs. Gauss Ratio (RMSE)} ($I_{\rm RMSE}$) &
        $I_{\rm RMSE} = \dfrac{\mathrm{RMSE}_{\rm Ekman}}{\mathrm{RMSE}_{\rm Gauss}}$ &
        Values $<1$ indicate the Ekman model outperforms the Gauss–Markov model; lower is better \\
        \hline
        \textit{Mean Copernicus Current Magnitude} ($\mu_{\rm Cop}$) &
        $\mu_{\rm Cop} = \dfrac{1}{N}\sum_{i=1}^N \sqrt{u_{\rm Cop,i}^2 + v_{\rm Cop,i}^2}$ &
        Provides a scale-independent normalisation factor for errors \\
        \hline
        \textit{Relative MAE} ($\mathrm{Rel\_MAE}_M$) &
        $\mathrm{Rel\_MAE}_M = \dfrac{\mathrm{MAE}_M}{\mu_{\rm Cop}}$ &
        Normalised MAE; lower values indicate better model performance \\
        \hline
        \textit{Relative RMSE} ($\mathrm{Rel\_RMSE}_M$) &
        $\mathrm{Rel\_RMSE}_M = \dfrac{\mathrm{RMSE}_M}{\mu_{\rm Cop}}$ &
        Normalised RMSE; lower values indicate better model performance \\
        \hline
    \end{tabularx}
\end{table*}

\textbf{Results:} Table~\ref{tab:ekman-metrics} presents a comparative assessment of the Ekman layer model and the Gauss–Markov model. Across all depths and time intervals, the Ekman-based approach consistently lowers MAE and RMSE, with overall improvements ranging from approximately 40\% to 85\%. The most pronounced gains are observed in subsurface layers (0–100 m), where error reductions frequently exceed 70\%. Mid-depth zones (100–500 m) exhibit the largest enhancements, with $I_{\mathrm{MAE}}$ and $I_{\mathrm{RMSE}}$ often surpassing 75\%, highlighting the model’s effectiveness in capturing subsurface current dynamics. Deeper layers also benefit, showing typical improvements above 60\%, while surface layer results are more variable, with moderate gains around 40–55\%.

Normalised error ratios further confirm that the Ekman model achieves lower relative MAE and RMSE compared to the Gauss–Markov approach. Overall, these findings indicate that incorporating Ekman dynamics enables more accurate representation of wind-driven and Coriolis-influenced currents, enhancing the simulation of large-scale underwater current dynamics.

\begin{table*}[htbp]
    \centering
    \caption{Comparison of error metrics between Ekman and Gauss–Markov models}
    \label{tab:ekman-metrics}
    \scriptsize
    \renewcommand{\arraystretch}{1.3}
    \begin{tabular}{llcccccc}
    \hline
     & 
    \textbf{Depth Range} & 
    $\bm{I_{\mathrm{MAE}}}$ & 
    $\bm{I_{\mathrm{RMSE}}}$ & 
     $\bm{\mathrm{Rel\_MAE}_{\mathrm{Ekman}}}$ & 
     $\bm{\mathrm{Rel\_MAE}_{\mathrm{Gauss}}}$ & 
     $\bm{\mathrm{Rel\_RMSE}_{\mathrm{Ekman}}}$ & 
     $\bm{\mathrm{Rel\_RMSE}_{\mathrm{Gauss}}}$ \\
    \hline
    4 Nov. 2023 & Surface (0–100 m)   & 41.18\% & 31.47\% & 1.52 & 2.59 & 1.87 & 2.74 \\
                    & Shallow (100–200 m) & 73.23\% & 71.11\% & 1.85 & 6.92 & 2.05 & 7.11 \\
                    & Mid-depth (200–500 m) & 74.93\% & 74.11\% & 2.14 & 8.54 & 2.27 & 8.78 \\
                    & Deep (500 m+)       & 80.29\% & 79.57\% & 2.03 & 10.30 & 2.16 & 10.57 \\
    \hline
    3 June 2024  & Surface (0–100 m)     & 44.73\% & 29.76\% & 1.20 & 2.18 & 1.63 & 2.32 \\
                 & Shallow (100–200 m)   & 79.26\% & 78.48\% & 0.47 & 2.27 & 0.52 & 2.42 \\
                 & Mid-depth (200–500 m) & 81.03\% & 79.25\% & 0.49 & 2.61 & 0.57 & 2.76 \\
                 & Deep (500 m+)         & 81.10\% & 79.72\% & 0.50 & 2.64 & 0.56 & 2.78 \\
    \hline
    31 July 2024 & Surface (0–100 m)     & 60.39\% & 59.71\% & 0.84 & 2.12 & 0.93 & 2.32 \\
                 & Shallow (100–200 m)   & 79.28\% & 78.45\% & 0.43 & 2.09 & 0.48 & 2.24 \\
                 & Mid-depth (200–500 m) & 76.85\% & 76.27\% & 0.61 & 2.65 & 0.68 & 2.85 \\
                 & Deep (500 m+)         & 40.57\% & 39.82\% & 0.79 & 1.32 & 0.90 & 1.49 \\
    \hline
    7 August 2024 & Surface (0–100 m)    & 55.37\% & 47.81\% & 1.07 & 2.40 & 1.30 & 2.49 \\
                  & Shallow (100–200 m)  & 71.30\% & 70.92\% & 0.96 & 3.34 & 1.03 & 3.55 \\
                  & Mid-depth (200–500 m) & 72.56\% & 72.28\% & 0.96 & 3.50 & 1.03 & 3.73 \\
                  & Deep (500 m+)        & 62.61\% & 61.07\% & 1.10 & 2.95 & 1.24 & 3.19 \\
    \hline
    3 Oct. 2024 & Surface (0–100 m)   & 46.40\% & 31.30\% & 0.93 & 1.74 & 1.27 & 1.86 \\
                   & Shallow (100–200 m) & 84.52\% & 83.66\% & 0.41 & 2.64 & 0.47 & 2.86 \\
                   & Mid-depth (200–500 m) & 80.03\% & 79.60\% & 0.58 & 2.88 & 0.65 & 3.16 \\
                   & Deep (500 m+)       & 82.88\% & 82.18\% & 0.44 & 2.55 & 0.50 & 2.82 \\
    \hline
    \end{tabular}

\end{table*}

\section{Features of the Simulator}

\subsection{Models}

LOTUSim provides open-source models, including:

\begin{itemize}
    \item Underwater vehicles: LRAUV, BlueROV
    \item Surface vessels: DTMB surface ship, PHAs, FREMM, Commando boat, WAM-V
    \item Aerial drones: X500 
    \item Environmental assets: mines, island, trees
    \item Human/robot avatars (Unity Animator handles motion and transitions)
\end{itemize}

\subsection{Robotics Sensors}
This subsection details the sensor suites integrated into the simulator, encompassing both onboard perception for autonomous systems and HITL sensing modalities. All sensor data are published as standard ROS2 messages.

For \textbf{surface navigation}, the simulator provides:
\begin{itemize}
    \item \textit{Inertial Measurement Unit (IMU)}: Provides high-frequency linear acceleration and angular velocity data, utilising the standard Gazebo sensor physics for realistic noise and bias modelling.
    \item \textit{2D LiDAR}: Delivers planar range sensing for obstacle detection and SLAM applications via the Gazebo ray-tracing plugin to simulate light-based distance measurements.
    \item \textit{Radar}: Simulates authentic maritime returns by convolving LiDAR point clouds with a range-dependent Point Spread Function (PSF) to create a rasterised display, following the methodology in \cite{lesy2025asvsimairsimsurfacevehicles}.
\end{itemize}

For \textbf{underwater navigation}, the suite includes:
\begin{itemize}
    \item \textit{IMU \& Magnetometer}: Provides high-frequency attitude estimation and heading reference data, utilising Gazebo's physics engine to simulate inertial and magnetic field vectors.
    \item \textit{Depth Sensor}: Measures the vehicle's vertical position using the global $z$-coordinate as a high-fidelity proxy for physical pressure sensors, consistent with modeling frameworks such as  \cite{romrell2025previewholoocean20,Player2023MultiAUV,Grimaldi2025}.
\end{itemize}

Additional \textbf{mission-specific} sensors and modules include:
\begin{itemize}
    \item \textit{Perception}: Leverages Unity-based RGB cameras for visual recognition and high-fidelity marine radar to facilitate long-range situational awareness in complex maritime environments.
    \item \textit{Operational Modules}: Integrates AIS (Automatic Identification System) data streams for vessel tracking and utilises a robust Waypoint Follower to enable autonomous path execution.
\end{itemize}

\subsection{HITL Sensing}
\begin{figure}[htbp]
    \centering
    \includegraphics[width=0.52\linewidth]{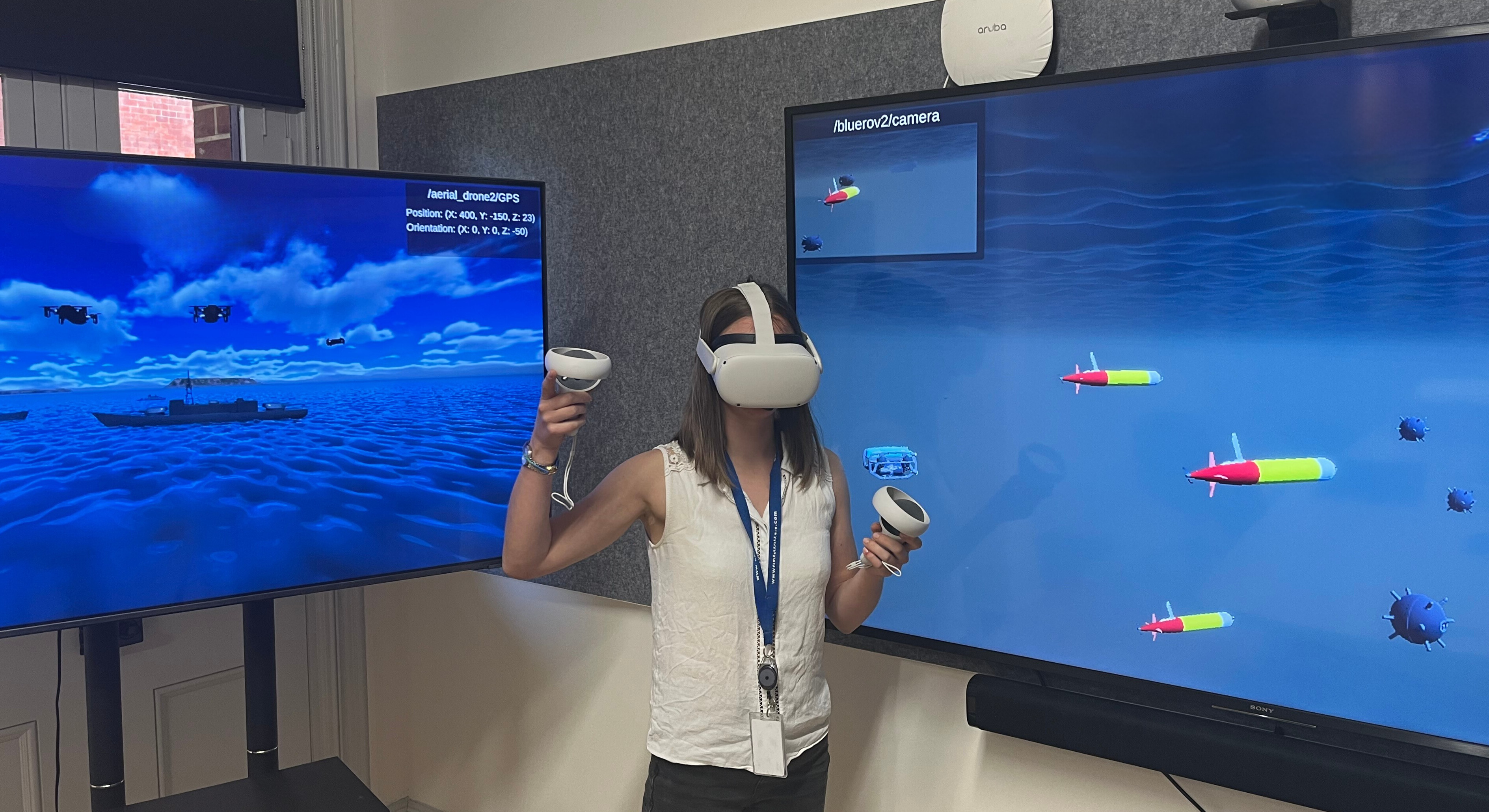}
    \includegraphics[width=0.46\linewidth]{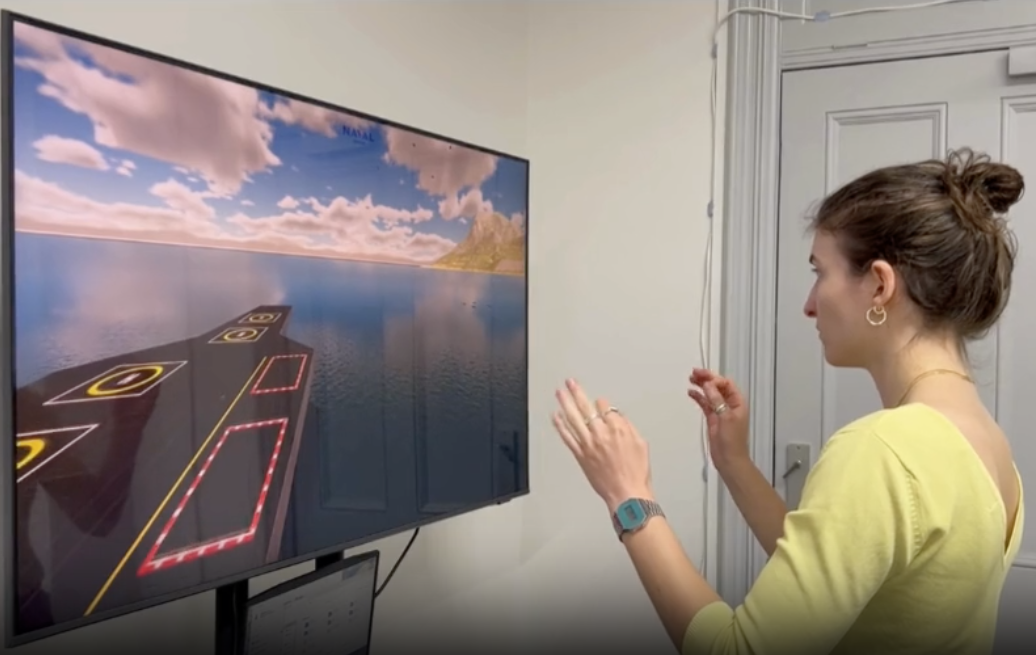}
    \caption{Left: user experimenting with VR in LOTUSim. Right: natural user-interaction interface.}
    \label{fig:operator-sensors}
\end{figure}

To enable seamless interaction between human operators and robotic agents, the simulator integrates sensors that capture human intent, motion, and cognitive state. This infrastructure allows users to control virtual entities or manage tactical interfaces through natural interaction modalities, as illustrated in Figure~\ref{fig:operator-sensors} and the accompanying video\footref{fn:video}.

\begin{itemize}
    \item \textit{Eye Tracker}: A non-immersive gaze-tracking system monitors fixation points and pupil metrics to infer user attention and assess physiological factors such as fatigue and cognitive load.
    \item \textit{Leap Motion}: A dedicated controller for high-fidelity hand and finger tracking, enabling gesture-based commands and precise human--machine interaction (HMI).
    \item \textit{VR}: Provides an immersive first-person perspective, tracking 6-DOF head motion to facilitate real-time human input and environmental feedback.
\end{itemize}

\section{Simulation-to-real}\label{sec:poc}

\begin{figure}[htbp]
    \centering
    \begin{tabular}{ccc}
        \includegraphics[width=0.25\linewidth]{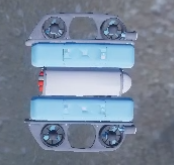} &
        \raisebox{0.4\height}{$\xrightarrow{\text{sim-to-real}}$} &
        \includegraphics[width=0.25\linewidth]{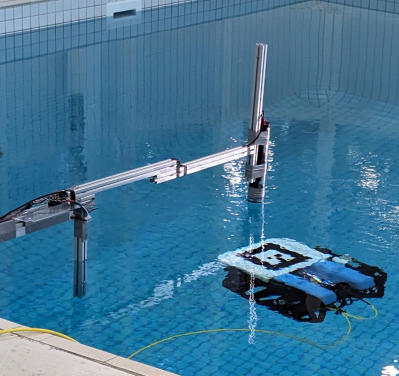}
    \end{tabular}
    \caption{Sim-to-real transfer for the BlueROV2 Heavy: (left) in LOTUSim; (right) in the test pool.}
    \label{fig:sim-to-real}
\end{figure}

In~\cite{lagattu2025control}, simulation results obtained with LOTUSim were successfully transferred to a physical BlueROV2 platform, demonstrating the simulator’s sim-to-real capability. In that study, a fault-tolerant control strategy for the BlueROV2 was first designed and evaluated in simulation. Subsequent real-world experiments conducted in a test pool (Figure~\ref{fig:sim-to-real})) validated the simulated behavior: the controller dynamically adapted motor commands in response to faults, without relying on explicit fault diagnosis.

\begin{figure}[htbp]
    \centering
        \includegraphics[width=0.7\linewidth]{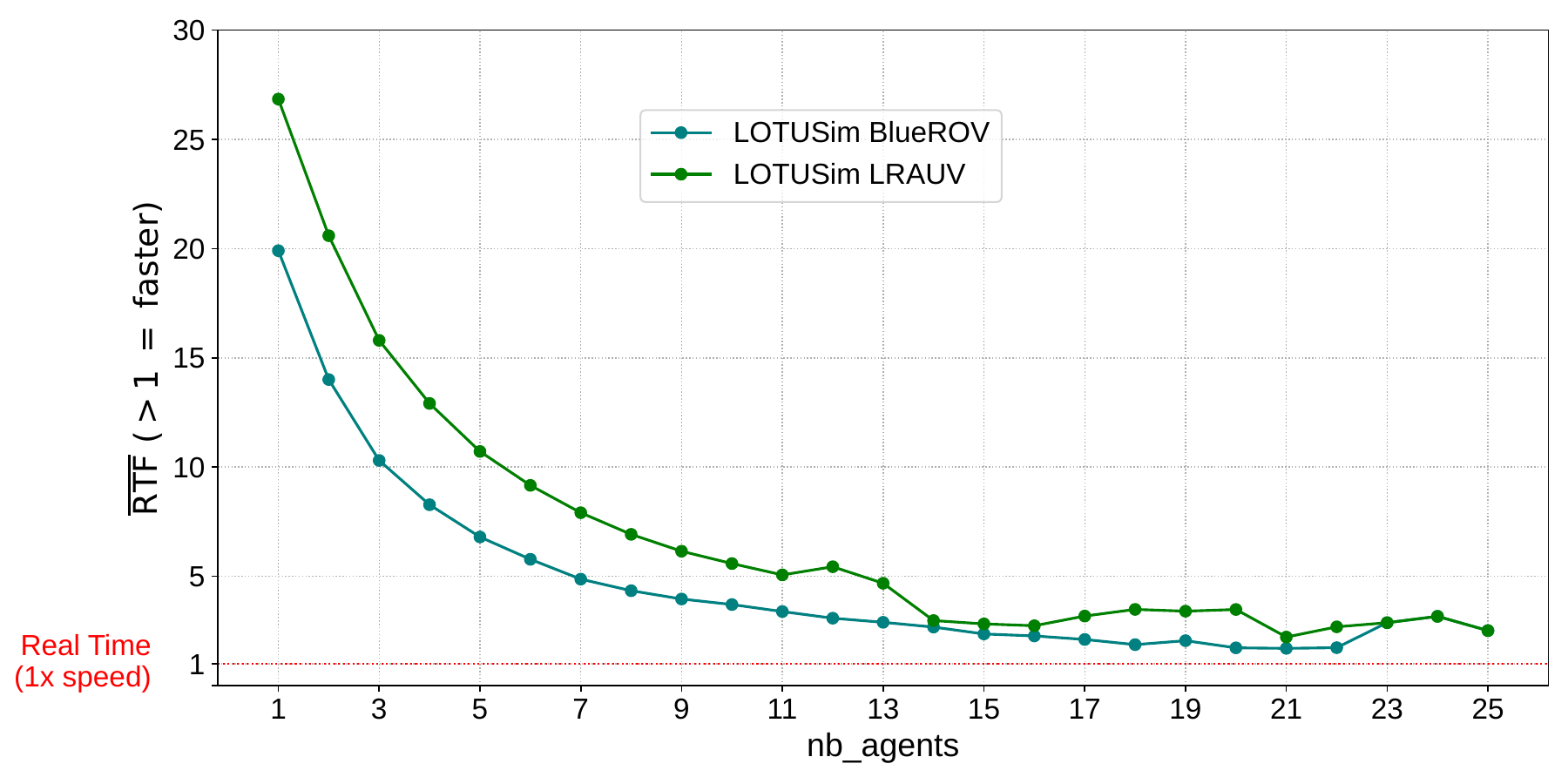}
        %\caption{RTF vs. number of agents (30ms)}
        \label{fig:lotusim_comparison_ai}
    
    \caption{RTF vs. number of agents (30ms)}
    \label{fig:RTF}
\end{figure}

In the present work, we exploit LOTUSim’s ability to accelerate the real-time factor (\texttt{RTF}) to enable efficient AI training. As illustrated in Figure~\ref{fig:RTF}, LOTUSim achieves an \texttt{RTF} exceeding 20 for a single drone. Even when scaling up to a swarm of 25 drones, the \texttt{RTF} remains above 1, allowing simulations to run faster than real-time.

%%%%%%%%%%%%%%%%%
\vspace{-1mm}
\section{Conclusion}
\label{sec:conclu}

\bibliographystyle{IEEEtran}
\bibliography{bib}  

\end{document}